\documentclass[11pt,a4paper,mathrsfs,mathserif]{article}
\usepackage[totalwidth=500pt,totalheight=700pt]{geometry}
\usepackage{graphicx,epstopdf,verbatim}
\usepackage[english]{babel}
\usepackage[unicode]{hyperref}
\usepackage{indentfirst}
\usepackage[width=0.8\textwidth,font=small]{caption}
\usepackage{amssymb}
\usepackage{amsmath}
\usepackage{authblk}
\usepackage[numbers,sort&compress]{natbib}
\begin{document}

\title{Scalar fields on $p$AdS}

\author{Feng Qu\thanks{qufeng@itp.ac.cn}}
\author{Yi-hong Gao\thanks{gaoyh@itp.ac.cn}}
\affil{\small{School of Physical Sciences, University of Chinese Academy of Sciences,\\No.19A Yuquan Road, Beijing 100049, China}}
\affil{\small{Key Laboratory of Theoretical Physics, Institute of Theoretical Physics, Chinese Academy of Sciences,\\Beijing 100190, China}}
\date{}

\maketitle

\begin{abstract}
We obtain a subgroup of the isometry group of $p$AdS (a $p$-adic version of AdS alternative to the Bruhat-Tits tree). We propose a candidate for the scalar bulk action and equation of motion on $p$AdS, and work out analytical expressions of the Green's functions for a particular choice of parameter together with an ansatz for general cases. The limiting behaviors of the Green's function are also studied. With their help, the convergence of small loops (whose radii are smaller than AdS length scale of $p$AdS) is analyzed.
\end{abstract}

\maketitle

\section{Introduction}
\label{sec:intro}

There are at least 2 reasons to study physics over $p$-adic numbers $\mathbb{Q}_p$. The first one is that all experimental data are rational numbers $\mathbb{Q}$, indicating that any field including $\mathbb{Q}$ is possibly used in physics. The second one is that the Archimedean property~\cite{Volovich2010,Volovich1987,Varadarajan2004} may not hold at small scales where the unknown theory of quantum gravity dominates. $\mathbb{Q}_p$ is a non-Archimedean field including $\mathbb{Q}$. It is widely used in physics ~\cite{Volovich2010,Volovich1987,Zelenov1994,Missarov1989,Freund1987,Vladimirov1989,Vladimirov1991,Smirnov1992,KOCHUBEI2003}.
The application of $\mathbb{Q}_p$ in the anti-de Sitter/conformal field theory correspondence (AdS/CFT)~\cite{Maldacena1998,Gubser1998,Witten1998} begins when the Bruhat-Tits tree (BTtree) is treated as
a $p$-adic version of AdS in~\cite{Gubser2017,Heydeman2016}. More properties of the BTtree are studied based on their work~\cite{Bhattacharyya2016,Bhattacharyya2018,Gubser2017b,Gubser2017a,Dutta2017}.

Significate difference between the BTtree and usual AdS exists: the holographic coordinate of the BTtree is discrete. To make it continuous, another $p$-adic version of AdS ($p$AdS) is introduced~\cite{Gubser2017}. Later on, one more such kind of spacetime is proposed~\cite{BhowmickRay2018} with a similar relation between bulk and boundary fields to that on the BTtree obtained. Our paper is devoted to further studies on $p$AdS. We study differences between $p$AdS and the BTtree, such as isometry group, the Green's function and Witten diagrams~\cite{Witten1998}. Section~\ref{sec:btandpads} gives introductions to $\mathbb{Q}_p$, scalar fields on the BTtree and $p$AdS spacetime. We also present a subgroup of the isometry group of $p$AdS. Section~\ref{sec:fieldonpads} is our main work containing the action, equation of motion (EOM) and the analytical Green's function of a scalar field on $p$AdS. The limiting behaviors of the Green's function and a critical parameter are also pointed out. Section~\ref{sec:smallloop} focuses on small loops in $p$AdS, which  are missing on the BTtree. We consider their convergence in this section. The last section is summary.

\section{The Bruhat-Tits tree and $p$AdS}
\label{sec:btandpads}

\begin{figure}
\centering
\includegraphics[width=0.75\textwidth]{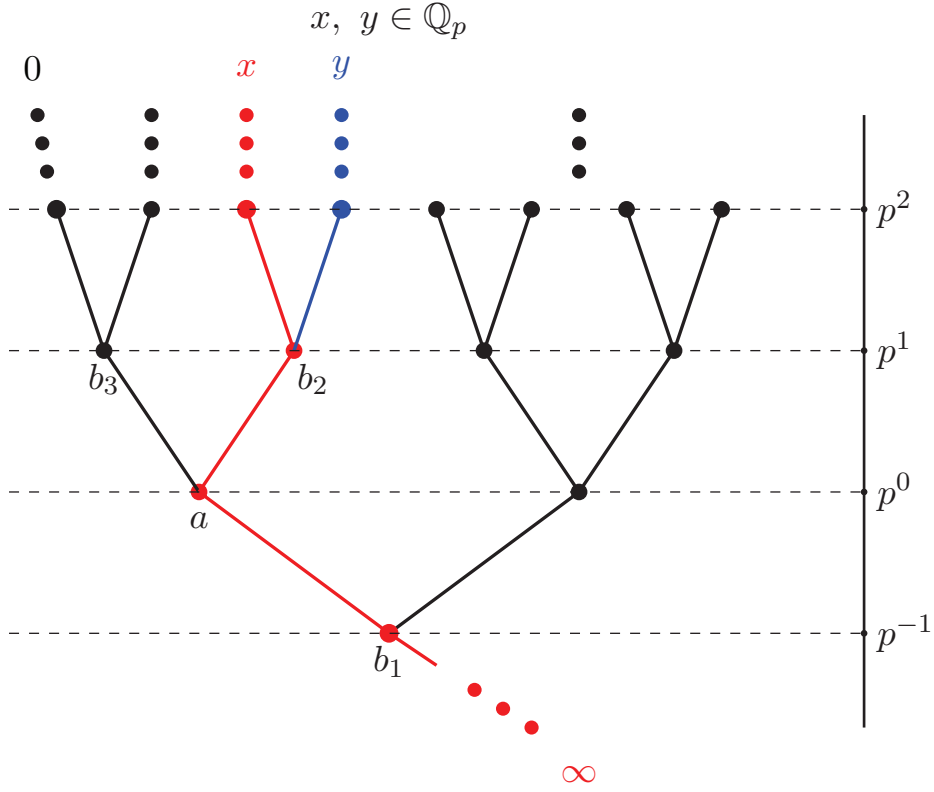}
\caption{\label{fig:bttree}The BTtree as a representation of $\mathbb{Q}_p$ ($p=2$). The vertical axis is place coordinate and its boundary (ends, or $\partial\textrm{BTtree}$) is $\partial\textrm{BTtree}=\mathbb{Q}_p\cup\{\infty\}$. $a$ and $b_i$ are 4 vertices (balls) satisfying $x,y\in b_2$ and $b_2+b_3=a\subset b_1$.}
\end{figure}

A non-zero $x\in\mathbb{Q}_p$ and its $p$-adic absolute value $|\cdot|$ read
\begin{equation}
x=\sum_{i=n}^{+\infty}a_ip^i\textrm{~where~}a_i\in\{0,1,\cdots,p-1\},a_n\neq0~.~|x|=p^{-n}~.
\end{equation}
$a_i$ is the digit at $p^i$ place. For $p=2$, letting $a_i=0(1)$ correspond to ``turn left (right)'' at the $i$-th step, walking from $p^{-\infty}$ to $p^{+\infty}$ place, a $2$-adic number can be represented by a broken line. In Fig.~\ref{fig:bttree}, the red line ($\infty\to\cdots\to b_1\to a\to b_2\to\cdots\to x$) represents $x=\cdots+0*2^{-1}+1*2^0+0*2^1+\cdots$.
$y$ is separated from $x$ at $2^1$ place: $|x-y|=|2^1|=2^{-1}$. The whole tree is the Bruhat-Tits tree (BTtree). Each vertex represents a ball containing every $p$-adic number whose line passes through it. If regarding vertices with place coordinate $p^1$ as the ends of the BTtree, balls $b_2,b_3,\cdots$ are identified as points in $\mathbb{Q}_p$. It gives a coarse-grained $\mathbb{Q}_p$. The deeper ``cutoff'' goes in Fig.~\ref{fig:bttree}, the larger balls treated as single points become. So the place coordinate can be regarded as the holographic dimension if identifying the BTtree as the AdS in AdS/CFT. Such $p$-adic AdS/CFT with the Euclidean time is built up in~\cite{Gubser2017}. The action, EOM of a scalar field on vertices $\phi_a$ with a point source and the Green's function depending only on the number of edges (spherical symmetry) are found to be
\begin{equation}
\label{eqn:bttreeeomgf}
\begin{aligned}
&S=\sum_{<ab>}\frac{1}{2}(\phi_a-\phi_b)^2+\sum_a\Big{(}\frac{1}{2}m_p^2\phi_a^2-\delta(a,a_0)\phi_a\Big{)}
\\
&\sum_{\substack{<ab>\\a\textrm{~fixed}}}(\phi_a-\phi_b)+m_p^2\phi_a=\delta(a,a_0)\equiv\begin{cases}
1~,~a=a_0
\\
0~,~a\neq a_0
\end{cases}
\\
&G(a,a_0)\equiv\phi_a=\frac{\zeta_p(2\Delta_p)}{p^{\Delta_p}}p^{-\Delta_p d(a,a_0)}\textrm{~where~}m_p^2=-\frac{1}{\zeta_p(\Delta_p-1)\zeta_p(-\Delta_p)}~.
\end{aligned}
\end{equation}
$<ab>$ means the sum is over the nearest neighboring vertices. $\zeta_p(s)\equiv\frac{1}{1-p^{-s}}$ and $d(a,a_0)$ gives the number of edges between $a$ and $a_0$.

\begin{figure}[tbp]
\centering
\includegraphics[width=0.8\textwidth]{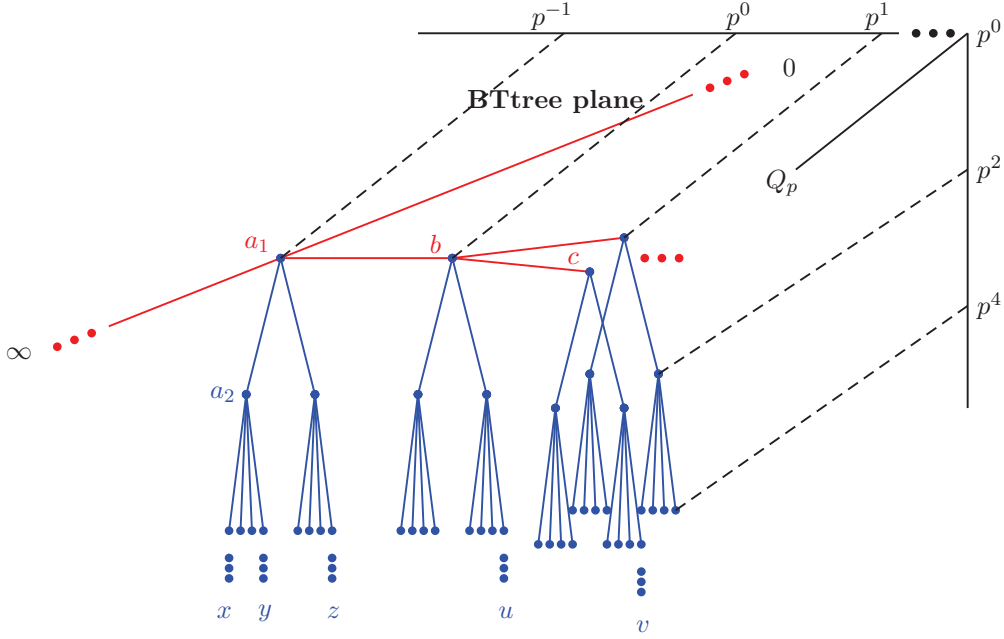}
\caption{\label{fig:pAdStree}$p$AdS represented by a tree containing red and blue edges ($p=2$). The vertical axis is level coordinate and the boundary of this tree is $p\textrm{AdS}\cup\partial p\textrm{AdS}$, where $\partial p\textrm{AdS}=\mathbb{Q}_p\cup\{\infty\}=\partial\textrm{BTtree}$. Red edges (the 1st level or the BTtree plane) show the same structure as that of the BTtree. $a_1,a_2,b$ and $c$ are 4 balls satisfying $x,y\in a_2\subset a_1,u\in b,v\in c$ and $a_1\cap b=a_1\cap c=b\cap c=\emptyset$. As for $u$ distance, we have $u(x,z)=|p^0|=p^{-0}, u(x,y)=|p^2|=p^{-2}, u(x,u)=p^{d(a_1,b)}=p^1$ and $u(x,v)=p^{d(a_1,c)}=p^2$.}
\end{figure}

$p\textrm{AdS}\equiv\mathbb{Q}_p^\times\times\mathbb{Q}_p$ ($\mathbb{Q}_p^\times\equiv\mathbb{Q}_p-\{0\}$) is introduced in~\cite{Gubser2017} equipped with a dimensionless distance $u(x,y)\equiv\frac{|x_1-y_1,x_2-y_2|_s^2}{|x_1y_1|}\textrm{~where~}
|x_1-y_1,x_2-y_2|_s=sup\{|x_1-y_1|,|x_2-y_2|\}$. Subscript $1$ denotes the holographic dimension. $p$AdS is represented by the tree in Fig.~\ref{fig:pAdStree}.
The ends of blue (red) lines in vertical (horizontal) direction make up $p$AdS ($\partial p\textrm{AdS}$). We call the vertical dimension where blue edges extend along ``level''. Each vertex represents a ball containing every point that has a blue line connecting it to the vertex. For points separated at the $n$-th level (still in the same ball at the 1st level, or in the same ``1st-level'' ball) with level coordinate $p^{2(n-1)}$,  $u$ distance between them (denoted by $u_n$) is $u_n=|p^{2(n-1)}|=p^{2(1-n)}$ which is always not larger than $1$. As for points belonging to different 1st-level balls, $u$ only depends on the number of red edges between these 1st-level balls. In this case $u$ is always larger than $1$. If treating points whose $u$ distance between them is not larger than $1$ as a single point, we can only recognize the structure of the 1st level. That means the BTtree can be obtained by coarse-graining $p$AdS. Measure $\mu$ is introduced as
\begin{equation}
\label{eqn:measure}
\mu_1\equiv\mu(a_1)\equiv\int_{x\in a_1}\frac{L_p^2d^2x}{|x_1|^2}=\frac{L_p^2}{\zeta_p(1)},\mu_{n+1}\equiv\mu(a_{n+1})=\frac{\mu_1p^{1-2n}}{p-1}~,
\end{equation}
where $a_n$ denotes a $n$-th-level ball. A length scale $L_p$ is introduced to make $\mu$ have the correct dimension. For series $\{u_n\}$ and $\{\mu_n\}$, we can summarise
\begin{equation}
u_n=p^{2(1-n)}~,~\mu_1=\frac{L_p^2}{\zeta_p(1)}~,~\mu_{n+1}=\frac{\mu_1p^{1-2n}}{p-1}~.
\end{equation}

The BTtree also can be regarded as a partition of $p$AdS under the equivalence relation ``$\sim$'': $x\sim y\Leftrightarrow u(x,y)\leq1$. Each 1st-level ball is an equivalence class. The transformation inside a 1st-level ball reads~\cite{Gubser2017}
\begin{equation}
\Lambda(\alpha,\beta)\equiv\begin{pmatrix}\alpha&0\\\beta&1\end{pmatrix}\textrm{~where~}\begin{cases}
\alpha\in\mathbb{U}_p=\{x|x\in\mathbb{Q}_p,|x|=1\}\\\beta\in\mathbb{Z}_p=\{x|x\in\mathbb{Q}_p,|x|\leq1\}\end{cases}~.
\end{equation}
$\Lambda$ acts on $x=(x_1,x_2)^\textrm{T}$ as the matrix multiplication. Supposing that representative elements have been chosen and fixed for all 1st-level balls, $\forall$ $x\in p\textrm{AdS}$ whose representative element is $x_{\textrm{rep}}$ we have $x=\Lambda_xx_{\textrm{rep}}$. The isometry group on the BTtree ``Isom(BTtree)'' can be regarded as a group acting on set $\{x_{\textrm{rep}}\}$. The action of Isom(BTtree) on $p$AdS can be defined as: $\forall g\in\textrm{Isom(BTtree)}$ and $\forall x\in p\textrm{AdS}$, $gx=g(\Lambda_xx_{\textrm{rep}})\equiv\Lambda_xgx_{\textrm{rep}}$. It can be verified that $\Lambda$ and $g$ are isometric transformations on $p$AdS (keep $u$ invariant) and commute with each other. So a subgroup of Isom($p$AdS) can be written as
\begin{equation}
\label{eqn:isom}
\textrm{Isom(BTtree)}\times\{\Lambda(\alpha,\beta)|\alpha\in\mathbb{U}_p,\beta\in\mathbb{Z}_p\}~.
\end{equation}
It can be verified that $\forall x\in\partial p\textrm{AdS},~\Lambda x=x$, which is a trivial action.

\section{Scalar fields on $p$AdS}
\label{sec:fieldonpads}

This section contains our main work. Using the correspondence between edges of a graph and the kinetic term of a field living on the same graph, we propose the action for a scalar field on $p$AdS in section~\ref{sub:actioneom} by refining the BTtree. In section~\ref{subsec:green}, we work out the analytical expressions of the Green's functions and point out the existence of a critical parameter.

\subsection{Action and EOM}
\label{sub:actioneom}

\begin{figure}[tbp]
\centering
\includegraphics[width=0.8\textwidth]{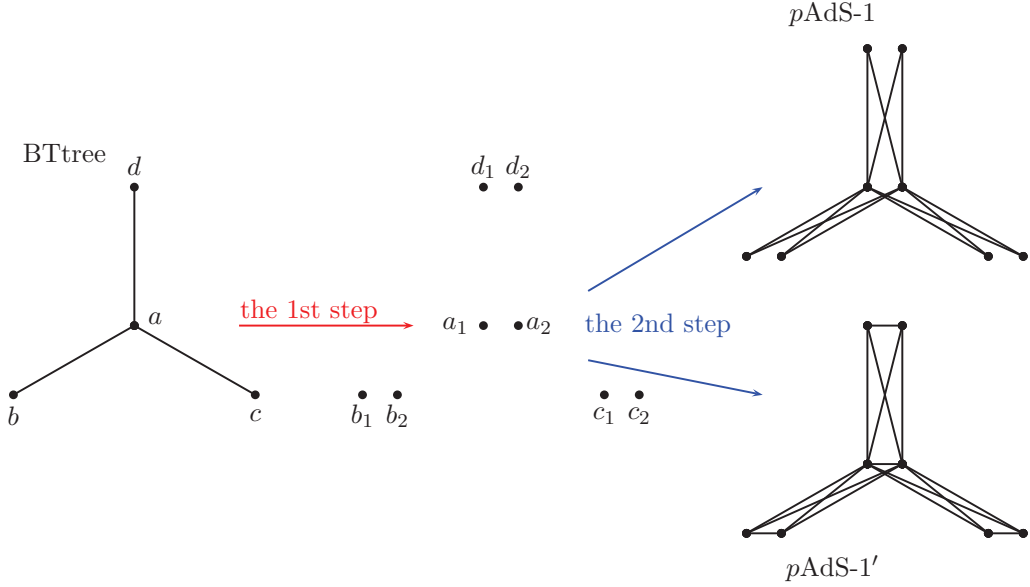}
\caption{\label{fig:pAdSn}2 steps for the first refining process of the BTtree. There are more short edges $a_1a_2,b_1b_2,c_1c_2$ and $d_1d_2$ in the graph $p$AdS-$1'$ compared with the graph $p$AdS-1.}
\end{figure}

Let $b,c,d$ denote the nearest neighboring vertices of $a$ on the BTtree (the left graph in Fig.~\ref{fig:pAdSn}).
Edges provide a natural representation of distances between vertices: the distance is determined by the number of edges connecting them. We'd like to go further to identify edges as the representation of the kinetic term. Specifically speaking, edge $ab$ gives a $\frac{1}{2}(\phi(a)-\phi(b))^2$ term, and the kinetic term is the sum of such terms over edges weighted by $d(a,b)^{-2}$. $d(a,b)$ denotes the length of $ab$. For example if setting $d=1$ for all edges, the BTtree (as a graph) gives the correct kinetic term in~\eqref{eqn:bttreeeomgf}.

$p$AdS can be obtained by refining the BTtree level by level (Fig.~\ref{fig:pAdStree}): decomposing each vertex at the 1st level into 2 vertices gives the 2nd level; decomposing each vertex at the 2nd level into 4 vertices gives the 3rd level and so on. To obtain the action on $p$AdS, firstly we need to obtain the graph representation of the $(n+1)$-th level (the graph ``$p$AdS-$n$'', where $n$ denotes the $n$-th refining process). Secondly write down the action using the above correspondence between edges and the kinetic term. Finally take the limit $n\to+\infty$.

Taking the first refining process for example, it requires 2 steps: decompose each vertex at the 1st level into 2 vertices and connect them with edges according to some rules. When adding edges, one rule we must obey is there should be edges between $a_i$ and $b_i,c_i,d_i$ since there are edges between $a$ and $b,c,d$ at the 1st level. Treating $i=1$ and $2$ equally, we add edges $a_ib_j$, $a_ic_j$ and $a_id_j$ to obtain a graph $p$AdS-$1$ (Fig.~\ref{fig:pAdSn}). It is a little strange that the distance between $a_1$ and $a_2$ is not represented by edges in this graph. If we demand that all distance information should be represented by edges, $p$AdS-$1$ is not the correct graph. We need to add short edges $a_1a_2$, $b_1b_2$, $c_1c_2$ and $d_1d_2$ to obtain another graph $p$AdS-$1'$. ``Short'' means $u(a_1,a_2)\equiv u(x,y)|_{x\in a_1,y\in a_2}=1=u(b_1,b_2)=u(c_1,c_2)=u(d_1,d_2)$. On the other hand, $a_ib_j$, $a_ic_j$ and $a_id_j$ are long edges satisfying $u(a_i,b_j)=u(a_i,c_j)=u(a_i,d_j)=p$. Repeating the same refining process, we can obtain the graph $p$AdS-$n'$, which is the correct graph representation of the $(n+1)$-th level in Fig.~\ref{fig:pAdStree}.

With $p$AdS-$n'$ in hand, we can write down the kinetic term for a scalar field living at the $(n+1)$-th level. Ignoring dimension problems, $p$AdS-$n'$ gives
\begin{equation}
\sum_{\langle a_{n+1}b_{n+1}\rangle}\frac{1}{2}\frac{(\phi(a_{n+1})-\phi(b_{n+1}))^2}{u(a_{n+1},b_{n+1})^\alpha}~.
\end{equation}
$\phi$ and $a(b)_{n+1}$ denote the scalar field and vertex at the $(n+1)$-th level in Fig.~\ref{fig:pAdStree}. The sum is over all edges of $p$AdS-$n'$, whose length ($u$ distance) equals to $p,1,p^{-2},p^{-4},\cdots,p^{-2(n-1)}$. Since $u$ is dimensionless, we introduce a parameter $\alpha>0$. Taking the limit $n\to+\infty$ will give the kinetic term of a field living on $p$AdS. Changing sum to integral with the measure~\eqref{eqn:measure}, the field theory on $p$AdS  is obtained as
\begin{equation}
\label{eqn:padseom}
\begin{aligned}
S=&\int dx\Big{(}\frac{1}{4}\int_{u(x,y)\leq p}dy\frac{(\phi(x)-\phi(y))^2}{L_p^4u(x,y)^\alpha}+\frac{1}{2}m^2\phi(x)^2-\delta(x,x_0)\phi(x)\Big{)}
\\
=&\int dx\Big{(}\frac{1}{2}\phi(x)\Box\phi(x)+\frac{1}{2}m^2\phi(x)^2-\delta(x,x_0)\phi(x)\Big{)}
\\
(\Box&+m^2)\phi(x)=\delta(x,x_0)~,
\end{aligned}
\end{equation}
where $dx\equiv\frac{L_p^2d^2x}{|x_1|^2}$, $\Box\phi(x)=\int_{u(x,y)\leq p}dy\frac{\phi(x)-\phi(y)}{L_p^4u(x,y)^\alpha}$. We set $L_p=1$ from now on. We can compare $\Box$ with the 2-dimension $s$-th-order Vladimirov operator~\cite{Vladimirov1991} $D^s$
\begin{equation}
\begin{aligned}
\Box\phi(x)=&\int_{u(x,y)\leq p}dy\frac{\phi(x)-\phi(y)}{|x_1-y_1,x_2-y_2|_s^{2\alpha}}|x_1y_1|^\alpha
\\
D^s\phi(x)=&\int_{{y\in(\mathbb{Q}_{p})^2}}d^2y\frac{\phi(x)-\phi(y)}{|x_1-y_1,x_2-y_2|_s^{2+s}}~.
\end{aligned}
\end{equation}
It seems $2\alpha\sim2+s$. We will talk about it at the ends of section~\ref{subsec:green} and~\ref{sec:smallloop}.

If we don't demand that edges should represent all distance information, $p$AdS-1(n) also can be used to construct the bulk action. Replacing the integral region $u(x,y)\leq p$ with $u(x,y)=p$ gives the result deduced from $p$AdS-$n$
\begin{equation}
S=\int dx\Big{(}\frac{1}{4}\int_{u(x,y)=p}dy\frac{(\phi(x)-\phi(y))^2}{L_p^4u(x,y)^\alpha}+\frac{1}{2}m^2\phi(x)^2-\delta(x,x_0)\phi(x)\Big{)}~,
\end{equation}
which is not considered in this paper.

\subsection{The Green's function and critical $\alpha$}
\label{subsec:green}

\begin{figure}[tbp]
\centering
\includegraphics[width=0.8\textwidth]{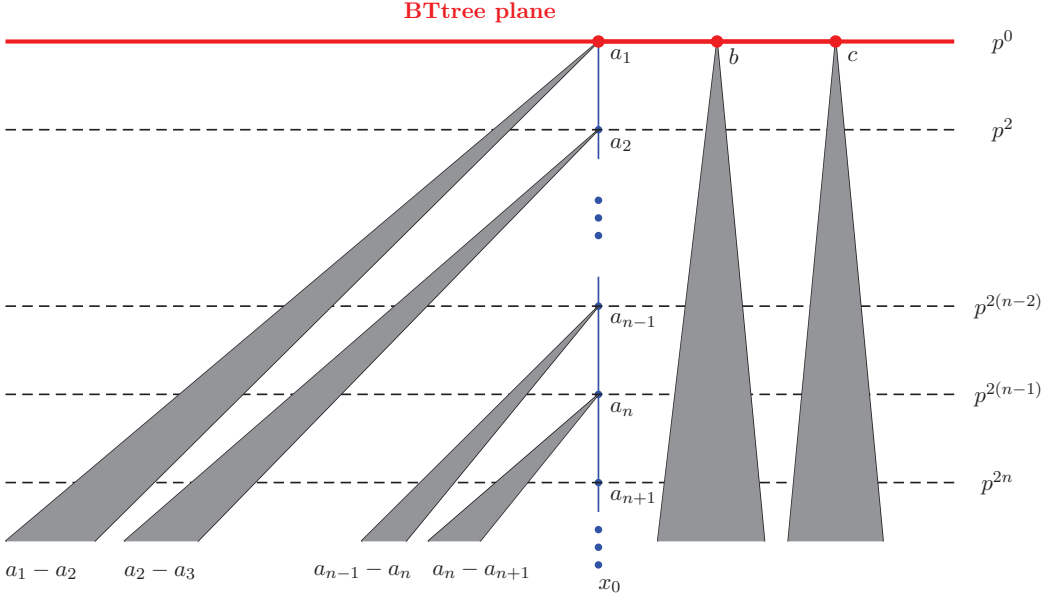}
\caption{\label{fig:Level}Spherical shells represented by shaded areas with point source $x_0$ at the center. $a_n$ is a $n$-th-level ball including $x_0$. The $u$ distance between any points in spherical shell $a_n-a_{n+1}$ and $x_0$ is $u=|p^{2(n-1)}|=p^{2(1-n)}$. $b$ and $c$ are two balls (spherical shells) those are 1 and 2 edges apart from $a_1$ at the 1st level.}
\end{figure}

Let $b_i$ denote the nearest neighboring vertex of $a$ at the 1st level. Integrating both sides of EOM~\eqref{eqn:padseom} with $\int_{x\in a}dx$ gives
\begin{equation}
\label{eqn:padstobttree}
\begin{aligned}
&\sum_i\Big{(}\frac{\mu_1}{p^\alpha}\phi(a)-\frac{\mu_1}{p^\alpha}\phi(b_i)\Big{)}+m^2\frac{p^\alpha}{\mu_1}\frac{\mu_1}{p^\alpha}\phi(a)
=\int_{x\in a}dx\delta(x,x_0)=\delta(a,a_1)
\\
&\textrm{where~}\phi(a)\equiv\int_{x\in a}dx\phi(x)~,~\phi(b_i)\equiv\int_{x\in b_i}dx\phi(x)\textrm{~and~}x_0\in a_1~.
\end{aligned}
\end{equation}
Referring to~\eqref{eqn:bttreeeomgf}, $\frac{\mu_1}{p^\alpha}\phi(a)$ can be regarded as a field on the BTtree with the mass square $m_p^2=m^2\frac{p^\alpha}{\mu_1}$, hence can be solved. Rewriting the parameter $\Delta_p$ in the solution of $\frac{\mu_1}{p^\alpha}\phi(a)$ as $\Delta\equiv\Delta_p$, we have $m^2\frac{p^\alpha}{\mu_1}=\frac{-1}{\zeta_p(\Delta-1)\zeta_p(-\Delta)}$. Since $u$ is discrete, other points form a series of spherical shells around $x_0$ (Fig.~\ref{fig:Level}).
Combining the solution of $\phi(a)$ and a spherical symmetry ansatz: $\phi(x)=f(u(x,x_0))$, the solution of EOM~\eqref{eqn:padseom} when $u(x,x_0)\geq p$ and $\int_{x\in a_1}dx\phi(x)$ read
\begin{equation}
\label{eqn:ugeqp}
\begin{aligned}
\phi(x)=&\frac{\zeta_p(2\Delta)}{\mu_1^2 p^{\Delta-\alpha}}p^{-\Delta d(x,x_0)}=\frac{\zeta_p(2\Delta)}{\mu_1^2 p^{\Delta-\alpha}}u(x,x_0)^{-\Delta}\textrm{~when~}u(x,x_0)\geq p
\\
\phi_1\equiv&\int_{x\in a_1}dx\phi(x)=\frac{\zeta_p(2\Delta)}{\mu_1 p^{\Delta-\alpha}}~.
\end{aligned}
\end{equation}

Set $\int_{x\in a_i}dx\phi(x)\equiv\phi_i$. When $x\in a_n-a_{n+1}$, $\phi(x)=\frac{\phi_n-\phi_{n+1}}{\mu_n-\mu_{n+1}}$. Integrating both sides of EOM \eqref{eqn:padseom} with $\int_{x\in a_n-a_{n+1}}dx$ and eliminating $\int_{u(x,y)=p}dy\frac{-\phi(y)}{u(x,y)^\alpha}$ with~\eqref{eqn:padstobttree}, we obtain
\begin{equation}
\label{eqn:rephi}
\frac{\phi_n-\phi_{n+1}}{\mu_n-\mu_{n+1}}=\frac{\sum_{i=1}^{n-1}(\frac{1}{u_i^\alpha}-\frac{1}{u_n^\alpha})(\phi_i-\phi_{i+1})+
(\frac{1}{u_n^\alpha}\phi_1+\frac{m^2}{\mu_1}\phi_1+\frac{p+1}{p^\alpha}\phi_1-\frac{1}{\mu_1})}
{\frac{\mu_n}{u_n^\alpha}+\frac{p+1}{p^\alpha}\mu_1+m^2+\sum_{i=1}^{n-1}\frac{\mu_i-\mu_{i+1}}{u_i^\alpha}}~.
\end{equation}
$n=1$ gives $\phi_2=(2-\frac{\mu_2}{\mu_1})\phi_1+\frac{\frac{\mu_2}{\mu_1}-1}{\frac{\mu_1}{u_1^\alpha}+\frac{p+1}{p^\alpha}\mu_1+m^2}$,
$n=2$ gives $\phi_3$ and so on. The analytic solution can be found for $\alpha=1,2\textrm{~and~}3$. Based on them we propose the ansatz
\begin{equation}
\label{eqn:alpha=n}
p^{2n}\phi_{n+1}=
\begin{cases}C_1+D_1\sum_{k=0}^{n-1}\frac{p^{2k}}{1-A_1k}&,\alpha=1
\\
C_\alpha+D_\alpha\sum_{k=0}^{n-1}\frac{p^{2k}}{1-A_\alpha p^{2(\alpha-1)k}}&,\alpha\in\mathbb{N}\textrm{~and~}\alpha\neq1\end{cases}~.
\end{equation}
$\mathbb{N}$ is the set of natural numbers. $n$-independent $C_\alpha,D_\alpha$ and $A_\alpha$ satisfy~\eqref{eqn:rephi}. This ansatz is confirmed for  $\alpha=1,2,3,4,5,6,10,20$ and $50$. We don't have a proof for general $\alpha\in\mathbb{N}$ or declaration for non-integral $\alpha$'s. The analytical expressions of $G(x,x_0)\equiv\phi(x)$ when $\alpha=2$ are summarized as
\begin{equation}
\label{eqn:alpha=2}
\begin{aligned}
&G(x,x_0)=\begin{cases}
\frac{\zeta_p(2\Delta)}{\mu_1^2p^{\Delta-2}}u(x,x_0)^{-\Delta}&,u\geq p
\\
\frac{\phi_1-\phi_2}{\mu_1-\mu_2}&,u=1
\\
\frac{p-1}{\mu_1p}\Big{(}C_2+D_2\sum_{k=0}^{n-1}\frac{p^{2k}}{1-A_2p^{2k}}-D_2\frac{1}{p^2-1}\frac{p^{2n}}{1-A_2p^{2n}}\Big{)}
&,u=p^{-2n}
\end{cases}
\\
&C_2=\frac{p^3+\phi_1(p-(1+m^2)p^4)}{(p-1)(1-(1+m^2)p^3)}~,~D_2=\frac{p^3-p^5}{1+p-m^2p^3}~,~A_2=\frac{p+p^3}{1+p-m^2p^3}~.
\end{aligned}
\end{equation}
$\phi_1(\phi_2)$ is the same as that in~\eqref{eqn:ugeqp}(the line below~\eqref{eqn:rephi}).

\begin{figure}[tbp]
\centering
\includegraphics[width=0.44\textwidth]{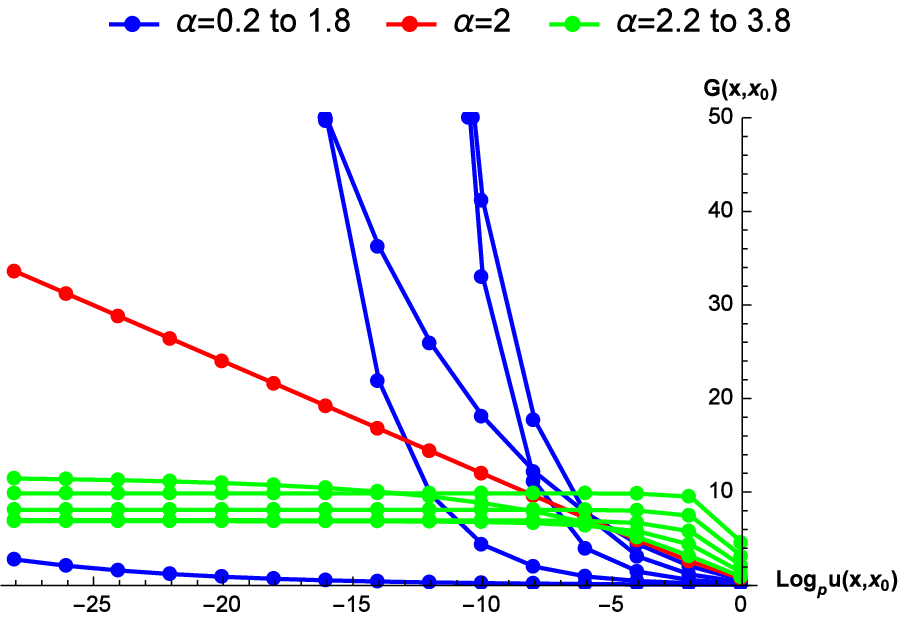}
\includegraphics[width=0.34\textwidth]{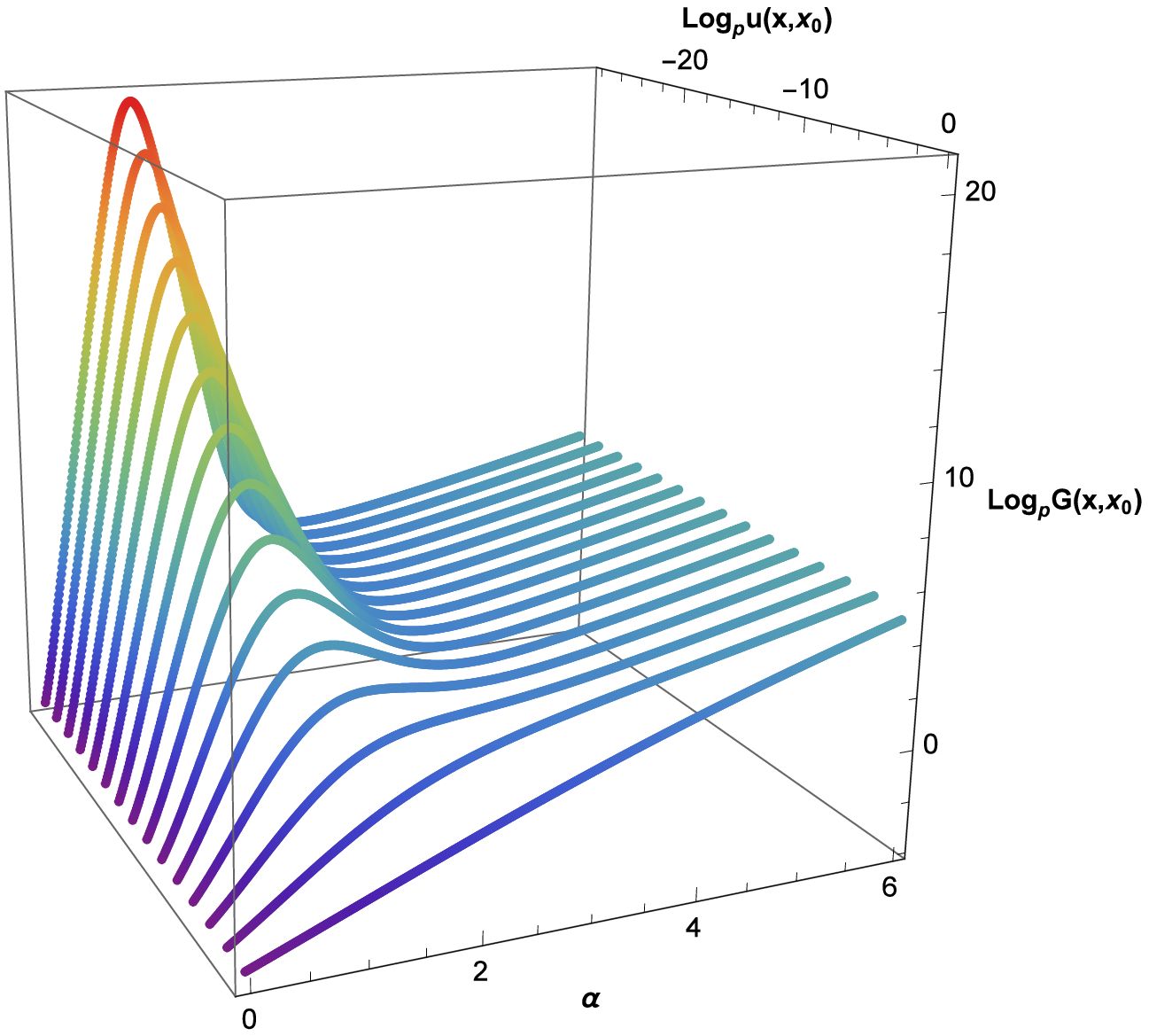}
\caption{\label{fig:plot}2D and 3D $G(x,x_0)$ versus $\log_pu(x,x_0)$ figures for different $\alpha$'s when $u(x,x_0)\leq1$. We use the $\phi_1$ in~\eqref{eqn:ugeqp} and set $p=2,\Delta=3$. The point source $x_0$ sits at $\log_pu(x,x_0)=-\infty$. In 2D figure, curves those go to $+\infty$ as $\log_pu\to-\infty$ correspond to cases of $0<\alpha\leq2$. Others those go to constants correspond to cases of $\alpha>2$. These numerical results give the critical value $\alpha=2$.}
\end{figure}

We numerically plot 2 figures in Fig.~\ref{fig:plot} for general $\alpha$'s.
The critical $\alpha=2$ can be confirmed analytically: after obtaining a recurrence relation $\phi_{n+3}=g_n\phi_{n+1}+h_n\phi_{n+2}$ from~\eqref{eqn:rephi},  $g_n(h_n)\xrightarrow[]{n\to+\infty}g(h)$ gives $\phi_{n+3}=g\phi_{n+1}+h\phi_{n+2}$, whose solution leads to
\begin{equation}
\label{eqn:limit}
G(x,x_0)\xrightarrow[]{x\to x_0}
\begin{cases}
c_{1}+c_{2}p^{2n}&,~0<\alpha<1
\\
c_{3}+c_{4}p^{2(2-\alpha)n}&,~\alpha\geq1\textrm{~and~}\alpha\neq2
\\
c_{5}+c_{6}n&,~\alpha=2~.
\end{cases}
\end{equation}
It gives the same critical $\alpha=2$. Remember the identification $2\alpha\sim2+s$ at the end of section~\ref{sub:actioneom}. $\alpha=2$ corresponds to the 2-dimension 2nd-order operator $D^2$ which leads to an expected EOM for a scalar field on a 2-dimension spacetime.

Different $\alpha$'s correspond to different theories. The $\alpha$-independent $G(x,x_0)$ when $\alpha<1$ indicates these theories have similar short-region behaviors. There may be some problems when we take the limit $n\to+\infty$ partly to obtain $\phi_{n+3}=g\phi_{n+1}+h\phi_{n+2}$. Consider 3 series $X_n=n,Y_n=1+\frac{1}{n}$ and $Z_n=-n$ satisfying $X_nY_n+Z_n=1$. Taking the limit partly leads to a wrong equation: $X_n\lim\limits_{n\to +\infty}Y_n+Z_n=1\Rightarrow 0=1$. The similar problem may exist here too, but we are not sure.

\section{Small loops in $p$AdS}
\label{sec:smallloop}

$p$-adic AdS/CFT also can be built up on $p$AdS, which leads to $p$AdS/CFT. The 2-point function can be calculated at tree level by $p$AdS/CFT using the same on-shell-action technique as that in~\cite{Gubser2017}. The only thing needed to be extra considered is the cutoff of $p$AdS: identify $x_1=$constant or $|x_1|=$constant as the boundary of $p$AdS. The former treats part of a 1st-level ball as the boundary, and the latter treats the whole 1st-level ball as the boundary. Using the latter cutoff, 2-point function at tree level by $p$AdS/CFT (with mass square $m^2$) differs from that of~\cite{Gubser2017} (with mass square $m_p^2=m^2\frac{p^\alpha}{\mu_1}$) only in the overall coefficient.

\begin{figure}[tbp]
\centering
\includegraphics[width=0.75\textwidth]{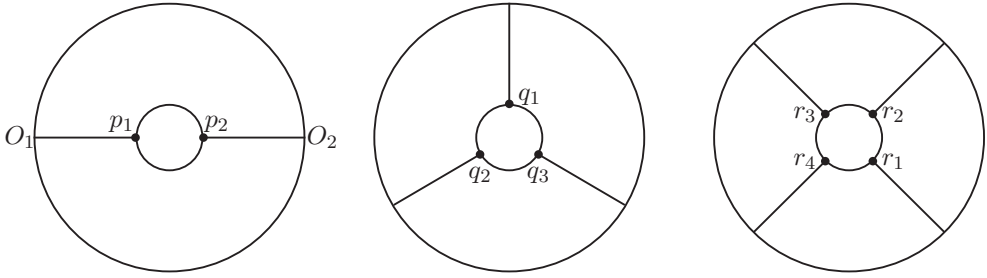}
\caption{\label{fig:witten}Small loops in $p$AdS. ``Small'' or ``radii are smaller than AdS length scale'' in the abstract means $u(p_1,p_2),u(q_i,q_j)$ and $u(r_i,r_j)\leq1$. Such small-loop diagrams are missing on the BTtree.}
\end{figure}

Considering that the Green's function of $p$AdS differs from that of the BTtree only in short region, the difference between these two spacetimes should show up in processes with fine structures, such as small-loop diagrams in Fig.~\ref{fig:witten}.
Actually they are Witten diagrams. Let $g,G,K$ denote the coupling constant, the Green's function and the bulk-boundary Green's function which is the regularization of $G$. For the left diagram in Fig.~\ref{fig:witten}, the corresponding amplitude is $g^2\int dp_1dp_2K(O_1,p_1)G(p_1,p_2)^2K(O_2,p_2)$. After simplification, the contribution of this small loop is represented by the factor $\int_{p_1\in a_1}dp_1G(p_1,p_2)^2=\sum_{n=1}^{+\infty}(\mu_n-\mu_{n+1})G(p_1,p_2)^2|_{u(p_1,p_2)=p^{2(1-n)}}$, where $a_1$ is a 1st-level ball. In large $n$ limit $\mu_n-\mu_{n+1}\propto p^{-2n}$. Combining with~\eqref{eqn:limit}, we can conclude that
\begin{equation}
\label{eqn:convergence}
\int_{p_1\in a_1}dp_1G(p_1,p_2)^2\textrm{~is~}\begin{cases}\textrm{~divergent},&0<\alpha\leq\frac{3}{2}\\\textrm{~convergent},&\alpha>\frac{3}{2}\end{cases}~.
\end{equation}
It is expected that no divergence is introduced by small loops when $\alpha>2$ since in such case $G(x,x_0)\xrightarrow{x\to x_0}\textrm{constant}$~\eqref{eqn:limit}. As for the left diagram in Fig.~\ref{fig:witten}, this lower limit can be lowed down to $\alpha>\frac{3}{2}$. It is worth mentioning that, according to the identification $2\alpha\sim2+s$ at the end of section~\ref{sub:actioneom}, these two critical values $\alpha=\frac{3}{2}$ and $\alpha=2$ give the upper and lower critical dimensions $s=1$ and $s=2$ respectively for a 2-dimension spacetime~\cite{GJPT2017}.

\section{Summary}
\label{sec:sum}

Based on~\cite{Gubser2017}, in this paper we (\romannumeral1)give a subgroup of Isom($p$AdS)~\eqref{eqn:isom}; (\romannumeral2)propose the action of a scalar field on $p$AdS~\eqref{eqn:padseom}; (\romannumeral3)work out the analytical expressions of the Green's functions for $\alpha=2$~\eqref{eqn:alpha=2} together with the ansatz for $\alpha\in\mathbb{N}$~\eqref{eqn:alpha=n} and their limiting behaviors~\eqref{eqn:limit}; (\romannumeral4)find out the critical value $\alpha=2$~\eqref{eqn:limit} and Fig.~\ref{fig:plot}; (\romannumeral5)point out that small loops in $p$AdS are missing on the BTtree, and analysis their convergence~\eqref{eqn:convergence}.

Some problems are still unknown, such as (\romannumeral1)can we embed $p$AdS in a higher dimension spacetime like~\cite{BhowmickRay2018} and~\cite{Guilloux2016}? (\romannumeral2)are there spinor or tensor fields in $p$AdS? (\romannumeral3)what is the universal lower limit of $\alpha$ for all small loops to be convergent?

\section*{Acknowledgement}

We are grateful to the referee for many valuable questions and suggestions.

This work is supported by the National Natural Science Foundation of China Grant No. 11747601 and 11875082.

\end{document}